\documentclass[aps,prl,twocolumn,showpacs]{revtex4}
\usepackage{bm}
\usepackage{amsmath}
\usepackage{amssymb}
\usepackage{graphicx}

\newcommand{\beq}{\begin{equation}}
\newcommand{\eeq}{\end{equation}}
\newcommand{\beqar}{\begin{eqnarray*}}
\newcommand{\eeqar}{\end{eqnarray*}}

\newcommand{\ETh}{E_{\text{Th}}}

\newcommand{\sea}{\searrow}
\newcommand{\nea}{\nearrow}
\newcommand{\swa}{\swarrow}
\newcommand{\nwa}{\nwarrow}

\newcommand{\Hh}{\hat{H}}

\newcommand{\rtarr}{\rightarrow}

\newcommand{\Hb}{{\bf H}}
\newcommand{\nb}{{\bf n}}
\newcommand{\tb}{{\bf t}}

\newcommand{\ib}{{\bf i}}
\newcommand{\jb}{{\bf j}}
\newcommand{\ab}{{\bf a}}
\newcommand{\bb}{{\bf b}}
\newcommand{\eb}{{\bf e}}

\newcommand{\s}{{\bf s}}

\newcommand{\qb}{{\bf q}}
\newcommand{\pb}{{\bf p}}

\newcommand{\Ab}{{\bf A}}
\newcommand{\rb}{{\bf r}}
\newcommand{\sbf}{{\bf s}}

\newcommand{\dg}{\dagger}

\newcommand{\lan}{\langle}
\newcommand{\ran}{\rangle}
\newcommand{\om}{\omega}
\newcommand{\Om}{\Omega}
\newcommand{\al}{\alpha}
\newcommand{\be}{\beta}
\newcommand{\ga}{\gamma}
\newcommand{\Ga}{\Gamma}
\newcommand{\de}{\delta}

\newcommand{\la}{\lambda}
\newcommand{\sig}{\sigma}
\newcommand{\Sig}{\Sigma}
\newcommand{\eps}{\varepsilon}

\newcommand{\lt}{\left}
\newcommand{\rt}{\right}
\begin{document}

\title{Hall resistivity of granular metals}
\author{M. Yu. Kharitonov$^{1}$ and K. B. Efetov$^{1,2}$}
\affiliation{$^{1}$ Theoretische Physik III, Ruhr-Universit\"{a}t Bochum, Germany,\\
$^{2}$L.D. Landau Institute for Theoretical Physics, Moscow, Russia.}
\date{\today}

\begin{abstract}

We calculate the Hall conductivity $\sig_{xy}$  and resistivity
$\rho_{xy}$ of a granular system at large tunneling conductance $g_{T}\gg 1$.
We show that in the absence of Coulomb interaction the Hall resistivity depends
neither on the tunneling conductance nor on the intragrain disorder
and is given by the classical
formula $\rho_{xy}=H/(n^* e c)$, where $n^*$ differs from the
carrier density $n$ inside the grains by a numerical coefficient
determined by the shape of the grains.
The Coulomb interaction gives rise to logarithmic in temperature
$T$ correction to $\rho_{xy}$ in the range $\Ga \lesssim T \lesssim \min(g_T E_c,\ETh)$,
where $\Ga$ is the tunneling escape rate, $E_c$ is the charging energy and
$\ETh$ is the Thouless energy of the grain.

\end{abstract}
\maketitle

Hall resistivity (HR) 
of metals and semiconductors gives very important
information about their properties.
According to the classical
Drude-Boltzmann theory HR
\begin{equation}
\rho _{xy}=H/(nec)
\label{eq:rxy}
\end{equation}
does not depend on the mean free path and allows one to
experimentally determine the carrier concentration $n$.

Recently, much attention from both experimental and theoretical sides has
been paid to granular systems (see a review \cite{BELVreview} and references therein).
Although various physical quantities have been calculated in
different regimes, the Hall transport in the granular matter has not been
addressed theoretically yet. In this work
we calculate the Hall conductivity(HC) of a granular system
and Coulomb interaction corrections to it
in the metallic regime, when the intergrain tunneling conductance
$G_T=(2e^2/\hbar) g_T$ is large, $g_T \gg 1$ (further we set $\hbar=1$).

Technically, calculating HC $\sig_{xy}$ appears
to be more complicated than calculating the longitudinal conductivity(LC) $\sig_{xx}$.
The granularity of the system is ensured by the condition that the conductance
$G_0=2e^2 g_0$ of the grain is much larger than the tunneling conductance $G_T$,
$g_0/g_T \gg 1$.
In this limit
the main contribution to $\sig_{xx}$ comes from the tunnel
barriers between the grains rather than from scattering on
impurities inside the grains. In the absence of Coulomb
interaction LC equals
\beq
    \sig^{(0)}_{xx}= G_T a^{2-d} ,
\label{eq:sigxx0}
\eeq
where $a$ is the size of the grains and $d$
is the dimensionality of the system.
Therefore when studying
longitudinal transport one can neglect electron dynamics inside
the grains, which simplifies calculations significantly.
On the contrary, for Hall transport one is forced to take
the intragrain electron dynamics into account,
since the Hall current
originates from the transversal drift
in crossed magnetic and electric fields {\em inside} the grains.

As we find in this work the intragrain electron dynamics
can be included within the diagrammatic approach
by considering
higher diffusion modes inside the grain.
This procedures accounts for the finiteness of the ratio $g_T/g_0$,
and allows one, in principle, to study
both LC and HC of the
granular system for arbitrary ratio $g_T/g_0$.
The obtained  results reproduce the
solution of the classical electrodynamics problem
for a granular medium (e.g. the formula $\sig^{(0)}_{xx}= a^{2-d} G_T G_0/(G_T+G_0)$
can be obtained as a series in $g_T/g_0$).
Quantum effects (e.g. Coulomb interaction and weak localization)
may be incorporated into this scheme afterwards.

We perform calculations for magnetic fields $H$ such that
$\om_H \tau \ll 1$,
where $\om_H = e H/mc$ is the Larmor frequency and $\tau$ is the  electron scattering time
inside the grain. Since the effective mean free path $l=v_F \tau \lesssim a$ does not
exceed the grain size $a$, and typically  $a \approx 10-100 \text{nm}$,
the condition $\om_H \tau \ll 1$ is well fulfilled for all experimentally available fields $H$.


Let us list the main results of this work.
First we neglect
Coulomb interaction completely and
obtain  for HC in the lowest nonvanishing order in $g_T/g_0$
\beq
    \sig^{(0)}_{xy}= G_T^2 R_H a^{2-d},
\label{eq:sigxy0}
\eeq
where $R_H$ is the classical Hall resistance of a single grain (see Fig.~1(right)).
This result obtained by diagrammatic methods is completely
classical provided the tunneling contact is viewed as a surface resistor with conductance $G_T$.
The HR of the system
\beq
    \rho^{(0)}_{xy} =
    R_H  a^{d-2}= H/(n^*_d e c)
\label{eq:rhoxy0}
\eeq
is given by the Hall resistance of a single
grain $R_H$, which depends on {\em the geometry of the grain}
but {\em not on the intragrain disorder}.
Eq.~(\ref{eq:rhoxy0}) defines the effective carrier density $n^*_d$ of
the granular medium. For a three-dimensional ($d=3$, many granular monolayers)
array $n^*_3= A n $
differs from the electron density $n$ in the grain
by a numerical factor $A$, $0<A\le 1$,
determined by the grain geometry
\footnote{Note that although the granular array may be two- ($d=2$)
or three-dimensional ($d=3$), the grains themselves are three-dimensional,
and $n$ is a three-dimensional density.}.
For grains of a simple geometry this factor
is given by the ratio of the largest cross section area $S$
to the  cross section area of the lattice cell $a^2$: $A=S/a^2$.
So, $A=1$  for cubic grains ($S=a^2$), $A=\pi/4$ for spherical grains ($S=\pi a^2/4$).
For a two-dimensional ($d=2$, single granular monolayer) array
the 3D result must be multiplied by the thickness of the layer $a$:
$n^*_2=  a A n $.


Next we calculated the first-order correction to HC
$\sig^{(0)}_{xy}$ (Eq.~(\ref{eq:sigxy0})) due to Coulomb
interaction at temperatures $T\gtrsim \Ga$ not much smaller than
the tunneling escape rate $\Ga=g_T \de$ ($\de$ is the mean level
spacing of the grain). We find significant $T$-dependent
corrections in the range $\Ga \lesssim T \lesssim g_T E_c$
($E_c=e^2/a$ is the charging energy of the grain), whereas for
$T\gtrsim g_T E_c$ the relative corrections are of the order of
$1/g_T$ or smaller \footnote{We do not consider weak localization
effects in this paper assuming they are suppressed by either
magnetic field or inelastic processes.}.

There exist two different contributions to HC
\beq
    \sig_{xy}=\sig^{(0)}_{xy}+\de\sig_{xy}^{(1)}+\de\sig_{xy}^{(2)}.
    \label{eq:sig}
\eeq
One of them, $\de\sig_{xy}^{(1)}$, can be attributed to
the renormalization of individual tunneling conductances $G_T$
({\em tunneling anomaly} \cite{AA,TA1,TA2}) in a granular medium
and has the form:
\beq
    \frac{\de\sig^{(1)}_{xy}}{\sig^{(0)}_{xy}} =-\frac{1}{\pi g_T d}
        \ln \frac{g_T E_c}{T}.
\label{eq:dsigxy1}
\eeq
The other one, $\de\sig_{xy}^{(2)}$, is due to {\em elastic cotunneling} (EC)\cite{EC},
it involves diffusion of an electron through the grain
(that is why it is suppressed at temperatures greater
than the Thouless energy of the grain $\ETh$):
\beq
    \frac{\de\sig^{(2)}_{xy}}{\sig^{(0)}_{xy}} =\frac{c_d}{4 \pi g_T }
        \ln \lt[\frac{\min(g_T E_c,\ETh)}{T}\rt].
\label{eq:dsigxy2}
\eeq
where $c_d$ is a numerical factor
Eq.~(\ref{eq:cd}).
Since the correction $\de\sig_{xy}^{(1)}$
merely renormalizes the conductance $G_T$ in
Eq.~(\ref{eq:sigxy0}), it does not affect the HR
$\rho_{xy}=\sig_{xy}/\sig^2_{xx}$. Indeed the correction
Eq.~(\ref{eq:dsigxy1}) is cancelled by the corresponding
correction to $\sig_{xx}$ \cite{ET,BELV,BELVreview}, describing
renormalization of $G_T$ in Eq.~({\ref{eq:sigxx0}). Therefore the
total correction $\de\rho_{xy}$ to HR
$\rho_{xy}=\rho_{xy}^{(0)}+\de\rho_{xy}$ is  due to the EC effect
(Eq.~(\ref{eq:dsigxy2})) only: \beq
    \frac{\de\rho_{xy}}{\rho_{xy}^{(0)}}=   \frac{\de\sig^{(2)}_{xy}}{\sig^{(0)}_{xy}}=
        \frac{c_d}{4 \pi g_T } \ln \lt[\frac{\min(g_T E_c,\ETh)}{T}\rt].
\label{eq:drhoxy}
\eeq

We conclude that the Hall resistivity $\rho_{xy}$ of a granular
metal at temperatures $T\gtrsim \min(g_T E_c,\ETh)$ is given by
Eq.~(\ref{eq:rhoxy0}) and is independent  of the intragrain and
tunnel contact disorder.
Measuring $\rho_{xy}$  at such $T$
and using Eq.~(\ref{eq:rhoxy0}) one can extract an important characteristic
of the granular system: its effective carrier density $n^*_d$.
At temperatures $\Ga \lesssim T \lesssim \min(g_T E_c,\ETh)$
Coulomb interaction leads to $\ln T$-dependent corrections to
$\rho_{xy}$. Comparison of Eqs.
(\ref{eq:rhoxy0})-(\ref{eq:drhoxy}) with experimental data may
serve as a good check of the theory developed here. The
logarithmic dependence $\rho_{xx}=a+b\ln T$ of granular metals has
been observed experimentally \cite{expRxx}, and $\rho_{xy}$ can
also be measured \cite{expRxy}. Our theory may also be applied to
indium tin oxide(ITO) materials (see e.g. \cite{ITO}). Another
related effect is the anomalous Hall effect in ferromagnetic
granular materials \cite{Mitra}.

\begin{figure}
\includegraphics[width=.48\textwidth]{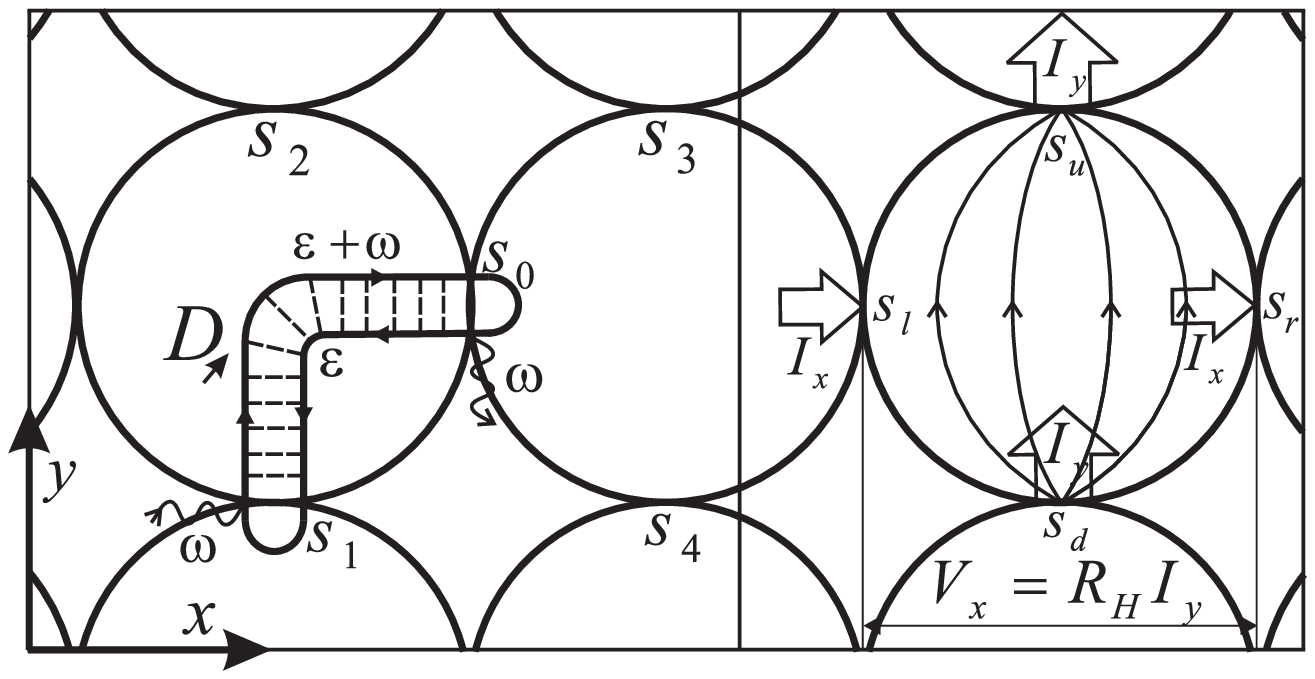}
\caption{ 
(left) Diagrams for the Hall conductivity
$\sig^{(0)}_{xy}$  of the granular system
(Eqs.~(\ref{eq:sigxy0D})). 
The diffuson $D_\nea$ connecting
contact $\sbf_1$ to $\sbf_0$ is shown. 
External tunneling vertices (wavy lines) must be attached in 4 possible ways.
Three more diffusons
$D_\sea, D_\swa, D_\nwa$ connecting contacts
$\sbf_2,\sbf_3,\sbf_4$ to $\sbf_0$, respectively,  must be also
taken into account.
(right) Classical picture of Hall conductivity
of a granular system. The current $I_y=G_T V_y$ running through
the grain in $y$-direction causes the Hall voltage drop $V_H=R_H
I_y$ between its opposite banks in $x$-direction, which is also
applied to the contacts (the total voltage drop per lattice period
in $x$-direction is 0) giving the Hall current $I_x=G_T V_H =
G_T^2 R_H V_y$ (Eq.~(\ref{eq:sigxy0})). }
\end{figure}

Now we briefly outline the model and method
used to derive the announced results, details of
our calculations will be presented elsewhere\cite{KEprep}.
We consider a quadratic($d=2$) or cubic($d=3$) lattice of identical (in form and size) grains
coupled to each other by tunnel contacts (Fig.~1).
To simplify the calculations we assume the intragrain electron dynamics diffusive,
$l \ll a$. However, our results
are also valid for ballistic ($l \sim a$)
intragrain disorder.
In the metallic regime ($g_T \gg 1$)
the Coulomb interaction effects
can be considered as a perturbation.


We write the Hamiltonian describing the system as
\[
    \Hh = \Hh_0 + \Hh_t + \Hh_c.
\]
\[
    \Hh_0= \sum_\ib \int d\rb_\ib \psi^\dg(\rb_\ib)
    \lt[ \xi\lt(\pb_\ib-e/c \Ab(\rb_\ib)\rt)+ U(\rb_\ib) \rt] \psi(\rb_\ib)
\]
is the electron Hamiltonian of isolated grains,
$\xi(\pb)=\pb^2/2m-\epsilon_F$,
$\Ab(\rb_\ib)$ is the vector potential describing uniform magnetic field $\Hb=H\eb_z$,
$U(\rb_\ib)$ is the random disorder potential of the grains,
$\ib=(i_1, \ldots , i_d)$ is an integer vector numerating the grains.
The disorder average is performed using a Gaussian distribution with the variance
$
    \lan U(\rb_\ib) U(\rb'_\ib) \ran = \frac{1}{2 \pi \nu \tau} \de(\rb_\ib-\rb'_\ib),
$
where $\nu$ is the density of states in the grain at the Fermi level per one spin direction.
Further, the tunneling Hamiltonian is given by
\[
    \Hh_t= \sum_{\lan \ib,\jb \ran} X_{\ib\jb}=
\sum_{\lan \ib,\jb \ran} \int
 d\sbf_\ib d\sbf_\jb \,
t(\sbf_\ib, \sbf_\jb) \psi^\dg(\sbf_\ib) \psi(\sbf_\jb)
\]
the summation is taken over the neighboring grains connected by a
tunnel contact, and the integration is done over the surfaces of
the contact. The inevitable irregularities of the tunnel contact
are modelled by the relation
$
    \lan t(\sbf_\ib, \sbf_\jb) t(\sbf_\jb, \sbf_\ib) \ran = t_0^2 \de(\sbf_\ib-\sbf_\jb),
$
where $\de(\sbf_\ib-\sbf_\jb)$ is a $\de$-function on the contact surface,
$t_0^2$ has the meaning of tunneling probability
per unit area of the contact.
The Coulomb interaction can be taken in the form (e.g. \cite{ABG})
\[
    \Hh_c= \frac{e^2}{2} \sum_{ \ib,\jb }
\frac{ n_\ib n_\jb }{C_{\ib\jb}},
\]
where $n_\ib=\int d\rb_\ib\, \psi^\dg(\rb_\ib) \psi(\rb_\ib) -\bar{n}$ is the excess number of electrons
in the $\ib$-th grain, $C_{\ib\jb}$  is the capacitance matrix.

The conductivity is calculated using the Kubo formula:
\beq
    \sig_{\ab\bb}(\om)= 2e^2 a^{2-d}\frac{1}{\om}
        \sum_\jb (\Pi_{\ab\bb}(\om, \ib-\jb)-\Pi_{\ab\bb}(0, \ib-\jb))
\label{eq:Sig}
\eeq
where $\om=2\pi T k >0 $ is a bosonic Matsubara frequency,
$\ab$ and $\bb$ are the lattice unit vectors,
\beq
    \Pi_{\ab\bb}(\om,\ib-\jb)=\int_0^{1/T} d\tau\,e^{i \om \tau} \lan T_\tau I_{\ib,\ab}(\tau) I_{\jb,\bb}(0) \ran
\label{eq:Pi}
\eeq
is the current-current correlator, $I_{\ib,\ab}(\tau) =X_{\ib+\ab,\ib}(\tau)-X_{\ib,\ib+\ab}(\tau)$,
the thermodynamic average $\lan\ldots\ran$ is taken with $\Hh$,
$A(\tau)=e^{\Hh\tau}A e^{-\Hh \tau}$ is the Heisenberg operator.

We perform calculations of $\sig_{\ab\bb}$ using diagrammatic technique.
Let us first neglect Coulomb interaction $\Hh_c$.
Technically, one expands Eq.~(\ref{eq:Pi})
both in disorder potential $U( \rb_\ib)$ and the tunnelling Hamiltonian $\Hh_t$
and averages over the intragrain and contact disorder.

Summing the diagrams corresponding to intragrain motion
we obtain an important object:
the diffusion propagator ({\em diffuson}) of a single isolated grain
given by a disorder-averaged product of two Green functions:
\[
    D(\om,\rb,\rb')=\frac{1}{2 \pi \nu} \lan G(\eps+\om,\rb,\rb')G(\eps,\rb',\rb)\ran,
            \mbox{ } (\eps+\om)\eps < 0.
\]
In the presence of magnetic field ($\om_H\tau \ll 1$)
this propagator
satisfies the equation
\beq
    (|\om|- D_0 \nabla_\rb^2) D(\om,\rb,\rb')=\de(\rb-\rb')
\label{eq:D}
\eeq
and the boundary condition at the grain surface
\beq
    (\nb, \nabla_\rb D)|_{\rb \in S}= \om_H  \tau (\tb, \nabla_\rb D)|_{\rb \in S}.
\label{eq:Dbc}
\eeq
Here $D_0=v_F l /3$ is the diffusion coefficient in the grain,
$\nb$ is the normal unit vector pointing outside the grain,
$\tb = [\Hb,\nb]/|[\Hb,\nb]|$ is the tangent unit vector
pointing in the direction of the edge drift.
Eq.~(\ref{eq:Dbc}) is due to the fact that the current component
normal to the grain surface vanishes,
the right hand side describes the edge drift caused by magnetic part of the Lorentz force.
The solution to Eqs.~(\ref{eq:D}),(\ref{eq:Dbc}) can be presented in the form
\beq
       D(\om, \rb,\rb')=
        \frac{1}{|\om| V}+ \sum_{n>0}
    \frac{ \phi_n(\rb) \phi^*_n(\rb')}{|\om|+\ga_n}
\label{eq:Dexpr}
\eeq
where $\phi_n$ are the eigenfunctions of the problem
\[ - D_0 \nabla_\rb^2 \phi_n=\ga_n \phi_n, \mbox{ }
    (\nb, \nabla_\rb \phi_n)|_S= \om_H  \tau (\tb, \nabla_\rb \phi_n)|_S.
\]
There always exists a uniform solution $\phi_0(\rb)=1/\sqrt{V}$
($V$ is the grain volume) with zero eigenvalue $\ga_0=0$
giving the 0-th harmonic $1/|\om| V$ in Eq.~(\ref{eq:Dexpr}).

Accounting for tunneling processes makes  the transport
between different grains possible.
The diagrams for Eq.~(\ref{eq:Pi})
in the $2N$-th order in $\Hh_t$
are given by diffusons connecting the contacts
$(\ib+\ab,\ib)$ and $(\jb+\bb,\jb)$ that run through
the other $N-1$ contacts in all possible ways.
An important observation is that the diffusons inside the grain
always enter the expression for $\sig_{\ab\bb}$ as a {\em difference}
$D(\om, \sbf_1,\sbf_2)-D(\om, \sbf_3,\sbf_4)$ of the diffusons connecting different contacts.
Therefore the 0-th harmonic
drops out and
the contribution to $\sig_{\ab\bb}$ comes from nonzero harmonics,
each 2 powers of $\Hh_t$ bringing a factor
$\Ga/\ETh=g_T/g_0$.

For the longitudinal conductivity ($\ab=\bb=\eb_x$)
the above procedure would lead to a small contribution
in the limit $g_T/g_0 \ll 1$, since Eq.~(\ref{eq:sigxx0})
can be obtained from a {\em single contact} ($\jb=\ib$)
without expanding Eq.~(\ref{eq:Pi})
in $\Hh_t$ ($N=0$).
For the Hall conductivity ($\ab=\eb_x, \bb=\eb_y$) considering the higher
 diffusion modes
is inevitable, because
even in the lowest nonvanishing order in $g_T/g_0$
we have to connect {\em different contacts} by diffusons of a single grain as shown
in Fig.~1(left) ($\jb=\ib$, $\ib+\eb_x$, $\ib-\eb_y$, $\ib+\eb_x-\eb_y$). Doing so, we obtain
\beq
    \sig^{(0)}_{xy}(\om)=  2 e^2 a^{2-d} \frac{g_T^2}{\nu}(D_\nea -D_\sea + D_\swa - D_\nwa),
\label{eq:sigxy0D}
\eeq
where $g_T=2 \pi (\nu t_0)^2 \Sig$ is the conductance of a tunnel contact,
$\Sig$ is the area of the contact,
$D_\al=\frac{1}{\Sig^2} \int d\s_0 d\s_a \bar{D}(\s_0,\s_a)$,
$a=1,2,3,4$ for $\al=\nea,\sea,\swa,\nwa$ respectively (Fig.~1(left)).
Here
\[  \bar{D}(\rb,\rb')=
        \sum_{n>0}
    \phi_n(\rb) \phi^*_n(\rb')/\ga_n
\]
is the diffuson without the 0-th harmonic at $\om=0$
satisfying Eqs.~(\ref{eq:D}),(\ref{eq:Dbc})  with $\om=0$
\footnote{ To get a correct $\om$-dependence of
Eq.~(\ref{eq:sigxy0D}) (its {\em independence} of $\om$) one
should renormalize the diffusons by the Coulomb interaction inside
the grain. This results in the replacement $1/(|\om|+\ga_n) \rtarr
1/\ga_n$ \cite{KEprep}. }.
Retaining only the 0-th harmonic in Eq.~(\ref{eq:Dexpr})
would give just $0$ in Eq.~(\ref{eq:sigxy0D}) and we are forced
to take all higher harmonics into account.
Eq.~(\ref{eq:sigxy0D}) is nonzero for $H\neq0$
since the edge trajectories for
$D_\nea  =D_\swa$ are shorter (if $e>0$ is assumed) than for $ D_\sea= D_\nwa$,
and therefore
$D_\nea - D_\sea=D_\swa - D_\nwa>0$.

In fact,
the result Eq.~(\ref{eq:sigxy0D}) is purely classical,
provided one  treats the tunnel contact as a surface resistor with the conductance $G_T$.
Indeed, the classical HC of the granular medium can be easily presented
in the form of Eq.~(\ref{eq:sigxy0}) (Fig.~2(right)).
The Hall resistance $R_H$
of the grain is
defined via the difference (Hall voltage) of
electric potential $\varphi(\rb)$ between the opposite banks of the grain,
$V_H=\varphi(\sbf_r)-\varphi(\sbf_l)=R_H I_y$,
when the current $I_y=I$ passes through the grain.
The current density $ \jb(\rb)=- \hat{\sig}_0 \nabla \varphi(\rb)$
($\hat{\sig}_0$ is the conductivity tensor) satisfies the continuity equation
$\text{div} \jb=q(\rb)$ and the boundary condition $(\nb\cdot \jb)|_S=0$.
The charge source function $q(\rb)$ is nonzero on the contacts surface only and
$\int d \sbf_d q(\sbf_d)=- \int d \sbf_u q(\sbf_u)=I$. Therefore $\varphi(\rb)$
satisfies
\beq
    -\nabla_\rb^2 \varphi = q(\rb)/\sig_{xx}^\text{gr},\mbox{ }
    (\nb, \nabla \varphi)|_S= \om_H  \tau (\tb, \nabla \varphi)|_S.
\label{eq:phi}
\eeq
Comparing Eq.~(\ref{eq:phi}) with Eqs. (\ref{eq:D}),(\ref{eq:Dbc})
we see that $ \bar{D}(\rb,\rb')$
is a Green function for the problem Eq. (\ref{eq:phi}).
Thus the solution to Eq.~(\ref{eq:phi}) is
\[
    \varphi(\rb)=\frac{1}{\nu} \frac{I}{\Sig}
        \lt( \int d\sbf_d \bar{D}(\rb,\sbf_d)-\int d\sbf_u \bar{D}(\rb,\sbf_u)\rt)
\]
and Eq.~(\ref{eq:sigxy0D}) leads to  Eq.~(\ref{eq:sigxy0})
(Einstein relation $\sig_{xx}^\text{gr}=2e^2 \nu D_0$ was used).
This establishes the connection between  our diagrammatic approach
of considering higher diffusion modes and the solution of the
classical electrodynamics  problem for the granular system.

Luckily, for simple geometries (cubic, spherical) of the grain the
Hall resistance $R_H$ can be obtained from symmetry arguments
without solving the problem Eq.~(\ref{eq:phi}). In such cases the
Hall voltage equals $V_H = \rho_{xy}^\text{gr} a I/S$, where $S$
is the area of the largest cross section of the grain,
$\rho_{xy}^\text{gr}=H/(nec)$ is the specific HR of the grain
material expressed in terms of the carrier density $n$ inside the
grain. Therefore, $R_H=\rho_{xy}^\text{gr}a/S$ and the HR of the
granular medium can be expressed in the form of
Eq.~(\ref{eq:rhoxy0}), where $n^*_d=a^{d-3} A n$, $A=S/a^2 \leq
1$. The quantity $n^*_d$ defines the effective carrier density of
the system.


Diagrammatic approach allows us to incorporate
quantum effects of Coulomb interaction
on Hall conductivity into the developed scheme.
We omit details
here leaving them for a more comprehensive
version \cite{KEprep}.
We consider the range of not very low temperatures $T\gtrsim \Ga$
where we can neglect the large scale contributions
analogous to those for homogeneously disordered metals \cite{AA}.
We find that in this regime in the first order in the screened Coulomb interaction
two contributions exist ($i=1,2$):
\[
    \de\sig^{(i)}_{xy}(\om)=  2 e^2 a^{2-d} \frac{g_T^2}{\nu}
        \sum_{n>0}(f_{n,\nea} -f_{n,\sea} + f_{n,\swa} - f_{n,\nwa}) \la_n^{(i)}(\om)
\]
where $f_{n,\al}=\frac{1}{\Sig^2} \int d\s_0 d\s_a \phi_n(\s_0)\phi_n^*(\s_a)$,
$a=1,2,3,4$ for $\al=\nea,\sea,\swa,\nwa$, respectively (Fig.~1(left)), and
\[
    \la^{(1)}_n(\om)=- 2 (2 \pi) T^2  {\sum_{\Om}}' 4
            \frac{V_0(\Om)-V_1(\Om)}{\ga_n \Om^2},
\]
\[
    \la^{(2)}_n(\om)= 2 (2 \pi) T^2 {\sum_{\Om}}'
        \frac{V_0(\Om)+V_2(\Om)-2 V_1(\Om)}{(\om+\Om+\ga_n)\Om^2}.
\]
Here $  2\pi T {\sum_\Om}' F(\Om) = \sum_{0<\Om \le \om} \Om  F(\Om)+
    \sum_{\om<\Om} \om  F(\Om)$,
$V_0=V(\ib,\ib)$, $V_1=V(\ib+\eb_x,\ib)$,
$V_2=V(\ib+\eb_x+\eb_y,\ib)$ are the components of the screened
Coulomb interaction $  V(\Om,\ib,\jb)=\int \frac{a^d
d^d\qb}{(2\pi)^d}    e^{i a \qb (\ib-\jb)} V(\Om,\qb)$ with
\[
    V(\Om,\qb)=E_c(\qb) /[1+2 E_c(\qb) \Ga_\qb /(\de|\Om|) ],
\]
$\Ga_\qb=2\Ga \sum_\be (1-\cos q_\be a)$, $E_c(\qb)=\sum_\ib e^{-i
a \qb (\ib-\jb)} e^2/C_{\ib-\jb}$ ($\be=x,y$ for $d=2$ and
$\be=x,y,z$ for $d=3$, $\Om=2\pi T m$, $m\in \mathbb{Z}$).
The
contribution $\de\sig_{xy}^{(1)}$ renormalizes individual
tunneling conductances
 $G_T$ in Eq.~(\ref{eq:sigxy0}),
whereas  the contribution $\de\sig_{xy}^{(2)}$ is
due to elastic cotunneling
(note the diffuson $1/(\om+\Om+\ga_n)$ of the grain in the expression for $\la_n^{(2)}$).
The large logarithms in $\la^{(1)}_n(\om)$ and $\la^{(2)}_n(\om)$
come from frequencies $\Om \lesssim g_T E_c$, for which the Coulomb potential
is  completely screened: $V(\Om,\qb)=\de |\Om|/2\Ga_\qb$.
Performing analytical continuation  and extracting $\sig_{xy}^{(0)}$ with
the help of Eq.~(\ref{eq:sigxy0D}) we arrive at Eqs.~(\ref{eq:dsigxy1}),(\ref{eq:dsigxy2})
with
\beq
    c_d=\int \frac{a^d d^d\qb}{(2\pi)^d} \frac{(1-\cos q_x a)(1- \cos q_y a)}
            {\sum_\be (1-\cos q_\be a)}.
\label{eq:cd}
\eeq

In conclusion, we presented a theory of the Hall conductivity of
granular metals. We have shown that at high enough temperatures
the Hall resistivity is given by a classical expression, from
which one can extract the effective carrier density of the system.
At lower temperatures charging effects give a logaritmic
 temperature dependence of Hall resistivity.
We hope that our predictions may be rather
easily checked experimentally.

Authors thank I. S. Beloborodov and A. F. Volkov for illuminating discussions
and acknowledge  financial support of Degussa AG (Germany),
SFB Transregio 12, the state of North-Rhine Westfalia, and European Union.


\begin{thebibliography}{99}

\bibitem{BELVreview} I. S. Beloborodov,  K. B. Efetov, A. V. Lopatin, and V. M. Vinokur,
    cond-mat/0603522, (2006).


\bibitem{AA} B.L. Altshuler and A.G. Aronov,
    in {\em Electron-Electron Interactions in Disordered Systems},
    edited by A.L. Efros and M. Pollak (Elsevier, Amsterdam, 1985).

\bibitem{TA1} Yu. V. Nazarov, Sov. Phys. JETP Lett. {\bf 49}, 126 (1989).

\bibitem{TA2} L. S. Levitov, A. V. Shytov, JETP Lett. {\bf 66}, 214 (1997).

\bibitem{EC} D. V. Averin and Yu. V. Nazarov, Phys. Rev. Lett. {\bf 65}, 2446 (1990).

\bibitem{ET} K. B. Efetov and A. Tschersich, Europhys. Lett. {\bf 59}, 114, (2002);
        Phys. Rev. B {\bf 67}, 174205, (2003).

\bibitem{BELV} I. S. Beloborodov, K. B. Efetov, A. V. Lopatin, and V. M. Vinokur,
        Phys. Rev. Lett {\bf 91}, 246801, (2003).

\bibitem{expRxx}
 A. Gerber, A. Milner, G. Deutscher, M. Karpovsky, and A. Gladkikh,
Phys. Rev. Lett. {\bf 78}, 4277, (1997);
Simon, R. W., B. J. Dalrymple, D. V. Vechten, W. W. Fuller, and S. A. Wolf,
Phys. Rev. B {\bf 36}, 1962, (1987)

\bibitem{expRxy} X. X. Zhang, Chuncheng Wan, H. Liu, Z. Q. Li, Ping Sheng, J. J. Lin,
            Phys. Rev. Lett. {\bf 86}, 5562 (2001);
    Peng Xiong, Gang Xiao, J. Q. Wang, John Q. Xiao, J. Samuel Jiang, C. L. Chien
 Phys. Rev. Lett. {\bf 69}, 3220 (1992).

\bibitem{ITO}
    J. Ederth {\em et al}, Phys. Rev. B {\bf 68}, 155410 (2003);
    J. Ederth {\em et al}, Thin Solid Films. {\bf 445}, 199 (2003).

\bibitem{Mitra}  P. Mitra, A. F. Hebard, K. A. Muttalib, P. Woelfle, cond-mat/0606215.

\bibitem{KEprep} M. Yu. Kharitonov, K. B. Efetov, in preparation.

\bibitem{ABG} I. L. Aleiner, P. W. Brouwer, and L. I. Glazman,
        Phys. Rep. {\bf 358}, 309 (2002).




\end{thebibliography}
\end{document}